\pdfoutput=1
\documentclass[12pt,a4paper]{article}

\usepackage{ifthen} % for conditional statements
\newboolean{pdflatex}
\setboolean{pdflatex}{true} % False for eps figures 

\newboolean{articletitles}
\setboolean{articletitles}{true} % False removes titles in references

\newboolean{uprightparticles}
\setboolean{uprightparticles}{false} %True for upright particle symbols

\usepackage{microtype}
\usepackage{lineno}  % for line numbering during review
\usepackage{xspace} % To avoid problems with missing or double spaces after
\usepackage{graphicx}  % to include figures (can also use other packages)
\usepackage{color}
\usepackage{colortbl}
\graphicspath{{./figs/}} % Make Latex search fig subdir for figures

\usepackage{amsmath} % Adds a large collection of math symbols
\usepackage{amssymb}
\usepackage{amsfonts}
\usepackage{bm}
\usepackage{upgreek} % Adds in support for greek letters in roman typeset

\newcommand*\patchAmsMathEnvironmentForLineno[1]{%
\expandafter\let\csname old#1\expandafter\endcsname\csname #1\endcsname
\expandafter\let\csname oldend#1\expandafter\endcsname\csname
end#1\endcsname
 \renewenvironment{#1}%
   {\linenomath\csname old#1\endcsname}%
   {\csname oldend#1\endcsname\endlinenomath}%
}
\newcommand*\patchBothAmsMathEnvironmentsForLineno[1]{%
  \patchAmsMathEnvironmentForLineno{#1}%
  \patchAmsMathEnvironmentForLineno{#1*}%
}
\AtBeginDocument{%
\patchBothAmsMathEnvironmentsForLineno{equation}%
\patchBothAmsMathEnvironmentsForLineno{align}%
\patchBothAmsMathEnvironmentsForLineno{flalign}%
\patchBothAmsMathEnvironmentsForLineno{alignat}%
\patchBothAmsMathEnvironmentsForLineno{gather}%
\patchBothAmsMathEnvironmentsForLineno{multline}%
}

\usepackage{hyperref}    % Hyperlinks in references
\usepackage[all]{hypcap} % Internal hyperlinks to floats.

\def\lhcb {\mbox{LHCb}\xspace}
\def\ux85 {\mbox{UX85}\xspace}

\ifthenelse{\boolean{uprightparticles}}%
{

 \def\PDelta      {\ensuremath{\Delta}\xspace}                 
 \def\PXi      {\ensuremath{\Xi}\xspace}                 
 \def\PLambda      {\ensuremath{\Lambda}\xspace}                 
 \def\PSigma      {\ensuremath{\Sigma}\xspace}                 
 \def\POmega      {\ensuremath{\Omega}\xspace}                 
 \def\PUpsilon      {\ensuremath{\Upsilon}\xspace}                 
 
 \def\PB      {\ensuremath{\mathrm{B}}\xspace}                 
                  
 \def\PD      {\ensuremath{\mathrm{D}}\xspace}

 \def\PK      {\ensuremath{\mathrm{K}}\xspace}

 \def\Pb      {\ensuremath{\mathrm{b}}\xspace}                 
 \def\Pc      {\ensuremath{\mathrm{c}}\xspace}

 \def\Pi      {\ensuremath{\mathrm{i}}\xspace}

 \def\Ps      {\ensuremath{\mathrm{s}}\xspace}

}
{

 \mathchardef\PDelta="7101
 \mathchardef\PXi="7104
 \mathchardef\PLambda="7103
 \mathchardef\PSigma="7106
 \mathchardef\POmega="710A
 \mathchardef\PUpsilon="7107
                  
 \def\PB      {\ensuremath{B}\xspace}                 
                  
 \def\PD      {\ensuremath{D}\xspace}

 \def\PK      {\ensuremath{K}\xspace}

 \def\Pb      {\ensuremath{b}\xspace}                 
 \def\Pc      {\ensuremath{c}\xspace}

 \def\Pi      {\ensuremath{i}\xspace}

 \def\Ps      {\ensuremath{s}\xspace}

}

\def\squark    {\ensuremath{\Ps}\xspace}

\def\cquark    {\ensuremath{\Pc}\xspace}

\def\bquark    {\ensuremath{\Pb}\xspace}

\def\kaon  {\ensuremath{\PK}\xspace}
  \def\Kbar  {\kern 0.2em\overline{\kern -0.2em \PK}{}\xspace}

\def\Kz    {\ensuremath{\kaon^0}\xspace}
\def\Kzb   {\ensuremath{\Kbar^0}\xspace}
\def\KzKzb {\ensuremath{\Kz \kern -0.16em \Kzb}\xspace}
\def\Kp    {\ensuremath{\kaon^+}\xspace}
\def\Km    {\ensuremath{\kaon^-}\xspace}

\def\KpKm  {\ensuremath{\Kp \kern -0.16em \Km}\xspace}

  \def\Dbar    {\kern 0.2em\overline{\kern -0.2em \PD}{}\xspace}
\def\D       {\ensuremath{\PD}\xspace}

\def\Dz      {\ensuremath{\D^0}\xspace}
\def\Dzb     {\ensuremath{\Dbar^0}\xspace}
\def\DzDzb   {\ensuremath{\Dz {\kern -0.16em \Dzb}}\xspace}
\def\Dp      {\ensuremath{\D^+}\xspace}
\def\Dm      {\ensuremath{\D^-}\xspace}

\def\DpDm    {\ensuremath{\Dp {\kern -0.16em \Dm}}\xspace}

\def\Dstarp  {\ensuremath{\D^{*+}}\xspace}

\def\B       {\ensuremath{\PB}\xspace}
  \def\Bbar    {\kern 0.18em\overline{\kern -0.18em \PB}{}\xspace}

\def\Bz      {\ensuremath{\B^0}\xspace}
\def\Bzb     {\ensuremath{\Bbar^0}\xspace}

\def\Bs      {\ensuremath{\B^0_\squark}\xspace}
\def\Bsb     {\ensuremath{\Bbar^0_\squark}\xspace}

  \def\Y#1S{\ensuremath{\PUpsilon{(#1S)}}\xspace}% no space before {...}!

\def\Lbar {\ensuremath{\kern 0.1em\overline{\kern -0.1em\PLambda}}\xspace}

\def\to                 {\ensuremath{\rightarrow}\xspace}

\def\order   {\ensuremath{\mathcal{O}}\xspace}

\def\CP                {\ensuremath{C\!P}\xspace}

\def\AT#1     {\ensuremath{A_{\mathrm{T}}^{#1}}\xspace}           % 2

\def\C#1      {\ensuremath{\mathcal{C}_{#1}}\xspace}                       % 9
\def\Cp#1     {\ensuremath{\mathcal{C}_{#1}^{'}}\xspace}                    % 7
\def\Ceff#1   {\ensuremath{\mathcal{C}_{#1}^{\mathrm{(eff)}}}\xspace}        % 9  
\def\Cpeff#1  {\ensuremath{\mathcal{C}_{#1}^{'\mathrm{(eff)}}}\xspace}       % 7
\def\Ope#1    {\ensuremath{\mathcal{O}_{#1}}\xspace}                       % 2
\def\Opep#1   {\ensuremath{\mathcal{O}_{#1}^{'}}\xspace}                    % 7

\newcommand{\tev}{\ensuremath{\mathrm{\,Te\kern -0.1em V}}\xspace}
\newcommand{\gev}{\ensuremath{\mathrm{\,Ge\kern -0.1em V}}\xspace}
\newcommand{\mev}{\ensuremath{\mathrm{\,Me\kern -0.1em V}}\xspace}
\newcommand{\kev}{\ensuremath{\mathrm{\,ke\kern -0.1em V}}\xspace}
\newcommand{\ev}{\ensuremath{\mathrm{\,e\kern -0.1em V}}\xspace}
\newcommand{\gevc}{\ensuremath{{\mathrm{\,Ge\kern -0.1em V\!/}c}}\xspace}
\newcommand{\mevc}{\ensuremath{{\mathrm{\,Me\kern -0.1em V\!/}c}}\xspace}
\newcommand{\gevcc}{\ensuremath{{\mathrm{\,Ge\kern -0.1em V\!/}c^2}}\xspace}
\newcommand{\gevgevcccc}{\ensuremath{{\mathrm{\,Ge\kern -0.1em V^2\!/}c^4}}\xspace}
\newcommand{\mevcc}{\ensuremath{{\mathrm{\,Me\kern -0.1em V\!/}c^2}}\xspace}

\def\invfb   {\ensuremath{\mbox{\,fb}^{-1}}\xspace}

\def\order{{\ensuremath{\cal O}}\xspace}

\def\gsim{{~\raise.15em\hbox{$>$}\kern-.85em
          \lower.35em\hbox{$\sim$}~}\xspace}
\def\lsim{{~\raise.15em\hbox{$<$}\kern-.85em
          \lower.35em\hbox{$\sim$}~}\xspace}

\def\tell1  {TELL1\xspace}
\def\ukl1   {UKL1\xspace}

\newcommand{\massgev}{\mbox{\gev/$c^2$}}
\newcommand{\massmev}{\mbox{\mev/$c^2$}}
\newcommand{\pgev}{\mbox{\gev/$c$}}
\newcommand{\pmev}{\mbox{\mev/$c$}}
\newcommand{\pis}{\ensuremath{\pi_{\rm s}}\xspace}
\newcommand{\M}{\ensuremath{M(\Dz\pis^+)}\xspace}
\usepackage{cite} % Allows for ranges in citations
\usepackage{mciteplus}

\begin{document}

%%%%%%%%%%%%%%%%%%%%%%%%%
%%%%% Title     %%%%%%%%%
%%%%%%%%%%%%%%%%%%%%%%%%%
\renewcommand{\thefootnote}{\fnsymbol{footnote}}
\setcounter{footnote}{1}
%%%%%%%%%%%%%%%%%%%%%%%%%
%%%%%  TITLE PAGE  %%%%%%
%%%%%%%%%%%%%%%%%%%%%%%%%
\begin{titlepage}
\pagenumbering{roman}

% Header ---------------------------------------------------
\vspace*{-1.5cm}
\centerline{\large EUROPEAN ORGANIZATION FOR NUCLEAR RESEARCH (CERN)}
\vspace*{1.5cm}
\hspace*{-0.5cm}
\begin{tabular*}{\linewidth}{lc@{\extracolsep{\fill}}r}
\ifthenelse{\boolean{pdflatex}}% Logo format choice
{\vspace*{-2.7cm}\mbox{\!\!\!\includegraphics[width=.14\textwidth]{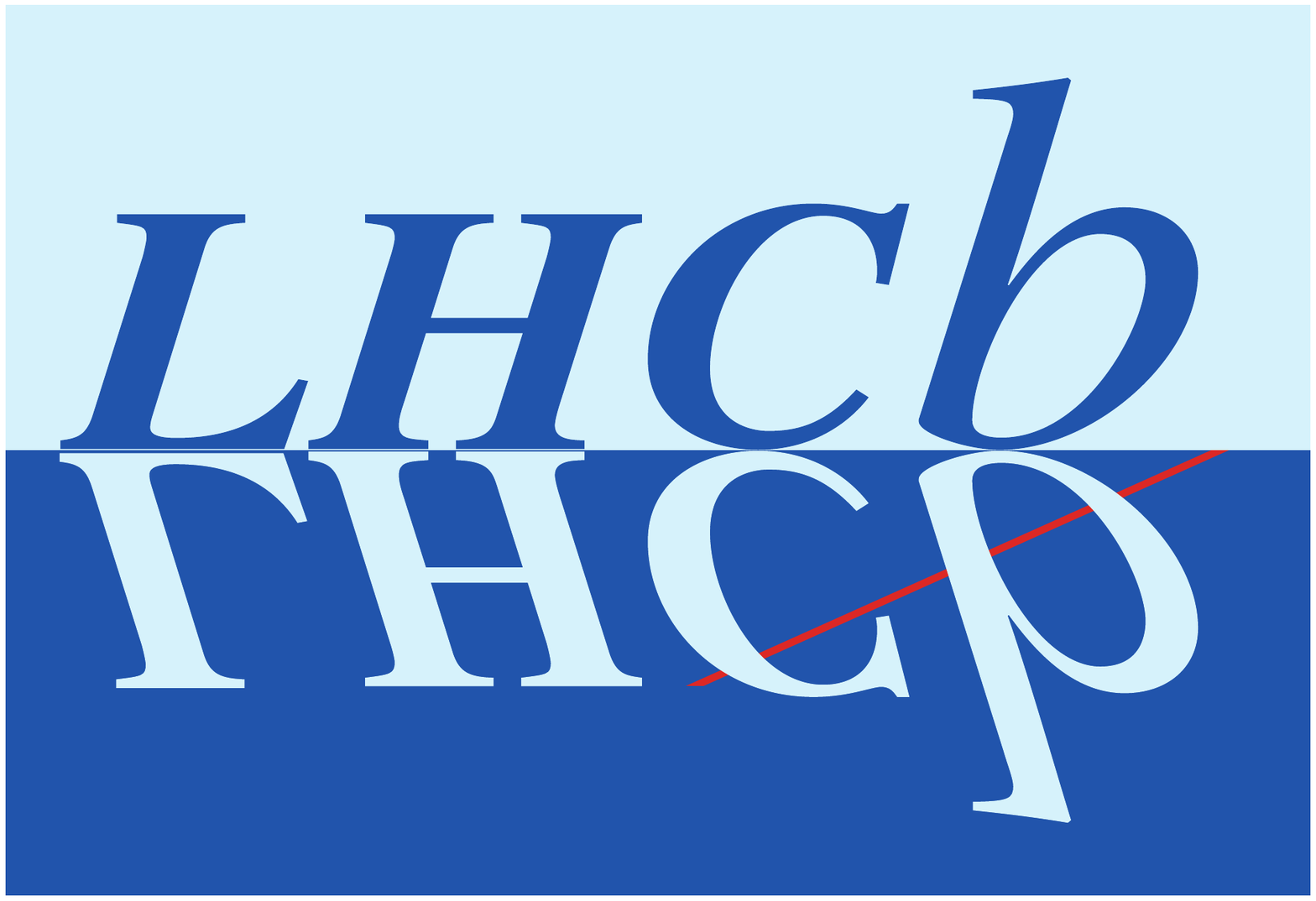}} & &}%
{\vspace*{-1.2cm}\mbox{\!\!\!\includegraphics[width=.12\textwidth]{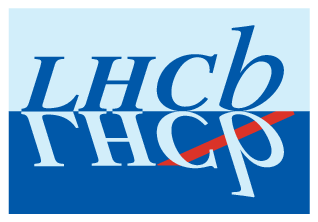}} & &}%
\\
 & & CERN-PH-EP-2012-333 \\  % ID 
 & & LHCb-PAPER-2012-038 \\  % ID 
 & & March 5, 2013 \\ % Date
 & & \\
\end{tabular*}

\vspace*{4.0cm}

% Title --------------------------------------------------
{\bf\boldmath\huge
\begin{center}
Observation of $\Dz-\Dzb$ oscillations
\end{center}
}

\vspace*{2.0cm}

% Authors -------------------------------------------------
\begin{center}
The LHCb collaboration\footnote{Authors are listed on the following pages.}
\end{center}

\vspace{\fill}

% Abstract -----------------------------------------------
\begin{abstract}
\noindent We report a measurement of the time-dependent ratio of $\Dz\to K^+\pi^-$ to $\Dz\to K^-\pi^+$ decay rates in \Dstarp-tagged events using $1.0\invfb$ of integrated luminosity recorded by the LHCb experiment. We measure the mixing parameters \mbox{$x'^2=(-0.9\pm1.3)\times10^{-4}$}, \mbox{$y'=(7.2\pm2.4)\times10^{-3}$} and the ratio of doubly-Cabibbo-suppressed to Cabibbo-favored decay rates \mbox{$R_D=(3.52\pm0.15)\times10^{-3}$}, where the uncertainties include statistical and systematic sources. The result excludes the no-mixing hypothesis with a probability corresponding to $9.1$ standard deviations and represents the first observation of $\Dz-\Dzb$ oscillations from a single measurement.
\end{abstract}

\vspace*{2.0cm}

\begin{center}
%Submitted to PRL
\end{center}

\vspace{\fill}

\end{titlepage}

%%%%%%%%%%%%%%%%%%%%%%%%%%%%%%%%
%%%%%  EOD OF TITLE PAGE  %%%%%%
%%%%%%%%%%%%%%%%%%%%%%%%%%%%%%%%

%  empty page follows the title page ----
\newpage
\setcounter{page}{2}
\mbox{~}
\newpage

% Author List ----------------------------
%%%%%%%%%%%%%%%%%%%%%%%%%%%%%%%%%%%%%%%%%%
\centerline{\large\bf LHCb collaboration}
\begin{flushleft}
\small
R.~Aaij$^{38}$, 
C.~Abellan~Beteta$^{33,n}$, 
A.~Adametz$^{11}$, 
B.~Adeva$^{34}$, 
M.~Adinolfi$^{43}$, 
C.~Adrover$^{6}$, 
A.~Affolder$^{49}$, 
Z.~Ajaltouni$^{5}$, 
J.~Albrecht$^{35}$, 
F.~Alessio$^{35}$
% ...

{\footnotesize \it
% ...
$ ^{5}$Clermont Universit\'{e}, Universit\'{e} Blaise Pascal, CNRS/IN2P3, LPC, Clermont-Ferrand, France\\
$ ^{6}$CPPM, Aix-Marseille Universit\'{e}, CNRS/IN2P3, Marseille, France\\
% ...
$ ^{11}$Physikalisches Institut, Ruprecht-Karls-Universit\"{a}t Heidelberg, Heidelberg, Germany\\
% ...
$ ^{33}$Universitat de Barcelona, Barcelona, Spain\\
$ ^{34}$Universidad de Santiago de Compostela, Santiago de Compostela, Spain\\
$ ^{35}$European Organization for Nuclear Research (CERN), Geneva, Switzerland\\
% ...
$ ^{38}$Nikhef National Institute for Subatomic Physics, Amsterdam, The Netherlands\\
% ...
$ ^{43}$H.H. Wills Physics Laboratory, University of Bristol, Bristol, United Kingdom\\
% ...
$ ^{49}$Oliver Lodge Laboratory, University of Liverpool, Liverpool, United Kingdom\\
% ...
$ ^{n}$LIFAELS, La Salle, Universitat Ramon Llull, Barcelona, Spain\\
% ...
}
\end{flushleft}
%%%%%%%%%%%%%%%%%%%%%%%%%%%%%%%%%%%%%%%%%%

\cleardoublepage

\renewcommand{\thefootnote}{\arabic{footnote}}
\setcounter{footnote}{0}

%%%%%%%%%%%%%%%%%%%%%%%%%
%%%%% Main text %%%%%%%%%
%%%%%%%%%%%%%%%%%%%%%%%%%
\pagestyle{plain} % restore page numbers for the main text
\setcounter{page}{1}
\pagenumbering{arabic}

Meson-antimeson oscillations are a manifestation of flavor changing neutral currents that occur because the flavor eigenstates differ from the physical mass eigenstates of the meson-antimeson system. Short-range quark-level transitions as well as long-range processes contribute to this phenomenon. The former are governed by loops in which virtual heavy particles are exchanged making the study of flavor oscillations an attractive area to search for physics beyond the standard model (SM). Oscillations have been observed in the $\Kz-\Kzb$ \cite{Lande:1956pf}, $\Bz-\Bzb$ \cite{Albrecht:1987dr} and $\Bs-\Bsb$ \cite{Abulencia:2006ze} systems, all with rates in agreement with SM expectations. Evidence of $\Dz-\Dzb$ oscillations has been reported by three experiments using different \Dz decay channels~\cite{Aubert:2007wf,Staric:2007dt,Aaltonen:2007ac,Aubert:2008zh,Aubert:2009ai}. Only the combination of these measurements provides confirmation of $\Dz-\Dzb$ oscillations, also referred to as charm mixing, with more than $5\sigma$ significance \cite{hfag}. While it is accepted that charm mixing occurs, a clear observation of the phenomenon from a single measurement is needed to establish it conclusively.

Charm mixing is characterized by two parameters: the mass and decay width differences, $\Delta m$ and $\Delta\Gamma$, between the two mass eigenstates expressed in terms of the dimensionless quantities $x = \Delta m/\Gamma$ and $y = \Delta\Gamma/2\Gamma$, where $\Gamma$ is the average $\Dz$ decay width. The charm mixing rate is expected to be small, with predicted values of $|x|,|y|\lesssim\mathcal{O}(10^{-2})$, including significant contributions from non-perturbative long-range processes that compete with the short-range electroweak loops \cite{Bianco:2003vb,Burdman:2003rs,Shipsey:2006zz,Artuso:2008vf}. This makes the mixing parameters difficult to calculate and complicates the unambiguous identification of potential non-SM contributions in the experimental measurements \cite{Petrov:2006nc,Golowich:2007ka,Ciuchini:2007cw}.

In the analysis described in this Letter, $\Dz-\Dzb$ oscillations are observed by studying the time-dependent ratio of $\Dz\to K^+\pi^-$ to $\Dz\to K^-\pi^+$ decay rates.\footnote{The inclusion of charge-conjugated modes is implied throughout this Letter.} The \Dz flavor at production time is determined using the charge of the soft (low-momentum) pion, $\pis^+$, in the strong $\Dstarp\to\Dz\pis^+$ decay. The $\Dstarp\to\Dz(\to K^-\pi^+)\pis^+$ process is referred to as right-sign (RS), whereas the $\Dstarp\to\Dz(\to K^+\pi^-)\pis^+$ is designated as wrong-sign (WS). The RS process is dominated by a Cabibbo-favored (CF) decay amplitude, whereas the WS amplitude includes contributions from both the doubly-Cabibbo-suppressed (DCS) $\Dz\to K^+\pi^-$ decay, as well as $\Dz-\Dzb$ mixing followed by the favored $\Dzb\to K^+\pi^-$ decay. In the limit of small mixing ($|x|,|y|\ll1$), and assuming negligible \CP violation, the time-dependent ratio, $R$, of WS to RS decay rates is approximated by \cite{Bianco:2003vb}
\begin{equation}\label{eq:true-ratio}
R(t) \approx R_D+\sqrt{R_D}\ y'\ \frac{t}{\tau}+\frac{x'^2+y'^2}{4}\left(\frac{t}{\tau}\right)^2,
\end{equation}
where $t/\tau$ is the decay time expressed in units of the average \Dz lifetime $\tau$, $R_D$ is the ratio of DCS to CF decay rates, $x' = x\cos\delta+y\sin\delta$, $y' = y\cos\delta-x\sin\delta$, and $\delta$ is the strong phase difference between the DCS and CF amplitudes.

The analysis is based on a data sample corresponding to $1.0\invfb$ of $\sqrt{s}=7\,\tev$ $pp$ collisions recorded by \lhcb during 2011. The \lhcb detector~\cite{Alves:2008zz} is a single-arm forward spectrometer covering the \mbox{pseudorapidity} range $2<\eta <5$, designed for the study of particles containing \bquark or \cquark quarks. Detector components particularly relevant for this analysis are the silicon Vertex Locator, which provides identification of displaced, secondary vertices of \bquark- and \cquark-hadron decays; the tracking system, which measures charged particles with momentum resolution $\Delta p/p$ that varies from $0.4\%$ at $5\,\pgev$ to $0.6\%$ at $100\,\pgev$, corresponding to a typical mass resolution of approximately $8\,\massmev$ for a two-body charm-meson decay; and the ring imaging Cherenkov detectors, which provide kaon-pion discrimination.

Events are triggered by signatures consistent with a hadronic charm decay. The hardware trigger demands a hadronic energy deposition with a transverse component of at least $3\,\gev$. Subsequent software-based triggers require two oppositely-charged tracks to form a \Dz candidate with a decay vertex well separated from the associated primary $pp$ collision vertex (PV). Additional requirements on the quality of the online-reconstructed tracks, their transverse momenta ($p_{\rm T}$) and their impact parameters (IP), defined as the distance of closest approach of the reconstructed trajectory to the PV, are applied in the final stage of the software trigger. For the offline analysis only \Dz candidates selected by this trigger algorithm are considered.

The \Dz daughter particles are both required to have $p_{\rm T}>800\,\pmev$, $p>5\,\pgev$ and $\chi^2(\text{IP})>9$. The $\chi^2(\text{IP})$ is defined as the difference between the $\chi^2$ of the PV reconstructed with and without the considered particle, and is a measure of consistency with the hypothesis that the particle originates from the PV. Selected \Dz candidates are required to have $p_{\rm T} > 3.5\,\pgev$ and are combined with a track with $p_{\rm T}>300\,\pmev$ and $p>1.5\,\pgev$ to form a \Dstarp candidate. Contamination from \D mesons originating from \bquark-hadron decays (secondary \D) is reduced by requiring the $\chi^2(\text{IP})$ of the \Dz and of $\pis^+$ candidates to be smaller than $9$ and $25$, respectively. In addition, the ring imaging Cherenkov system is used to distinguish between pions and kaons and to suppress the contamination from misidentified two-body charm decays in the sample. Backgrounds from misidentified singly Cabibbo-suppressed decays are specifically removed by requiring the \Dz candidate mass reconstructed under the $K^+K^-$ and $\pi^+\pi^-$ hypotheses to differ by more than $40\,\massmev$ from the known \Dz mass \cite{PDG2012}. Contamination from electrons to the soft pion sample is also suppressed using particle identification information. Finally, it is required that the \Dz and the $\pis^+$ form a vertex, which is constrained to the measured PV. Only candidates with reconstructed $K\pi$ mass within $24\,\massmev$ of the known \Dz mass and with reconstructed $\Dz\pis^+$ mass below $2.02\,\massgev$ are considered further. The $\Dz\pis^+$ mass, \M, is calculated using the vector sum of the momenta of the three charged particles and the known \Dz and $\pi^+$ masses \cite{PDG2012}; no mass hypotheses for the \Dz daughters enter the calculation, ensuring that all two-body signal decays have the same \M distribution \cite{Aaltonen:2011se}. Events with multiple RS or WS \Dstarp candidates occur about $2.5\%$ of the time, and all candidates are kept.

\begin{figure*}[t]
\centering
\includegraphics[width=0.45\textwidth]{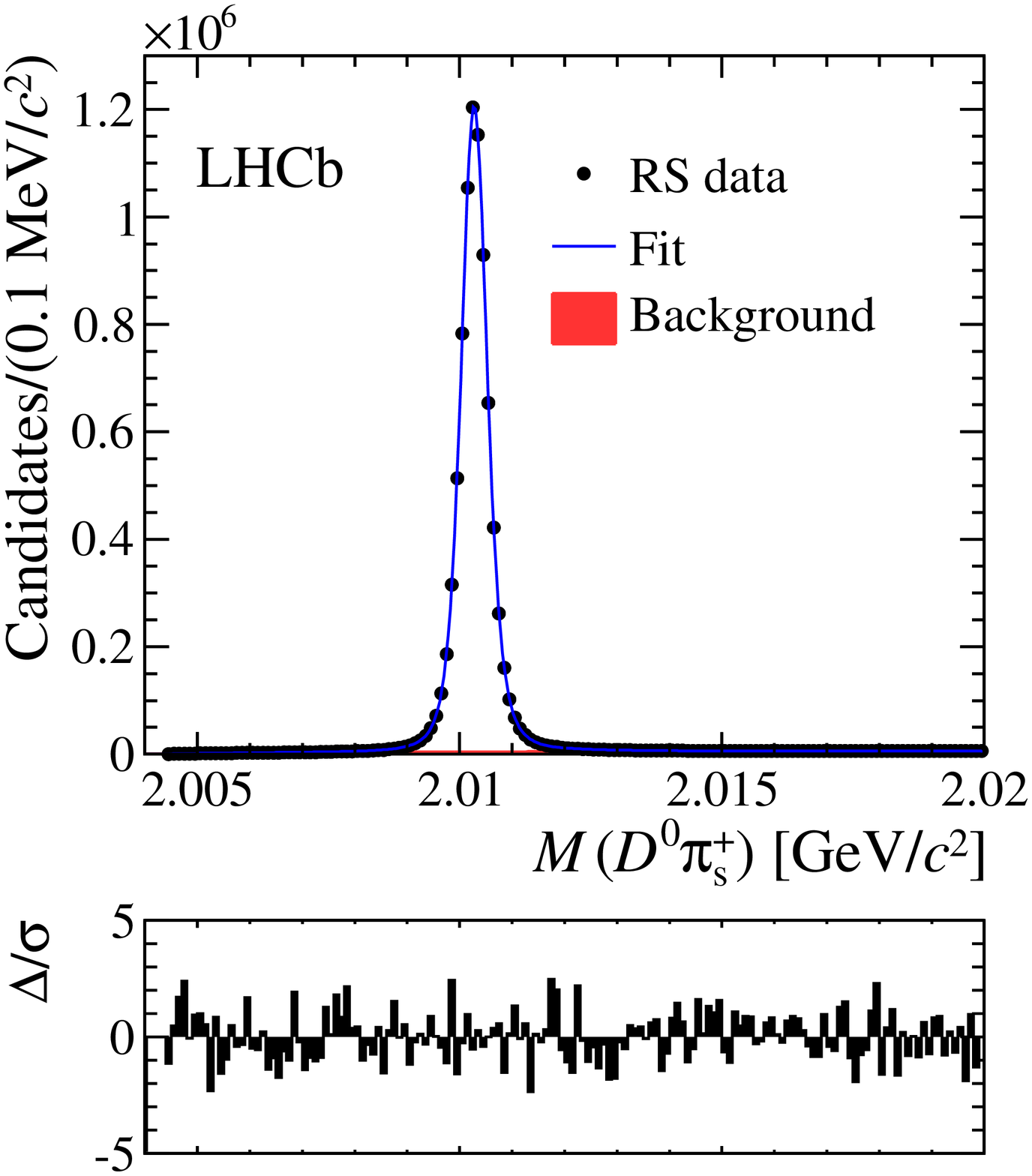}\hfil
\includegraphics[width=0.45\textwidth]{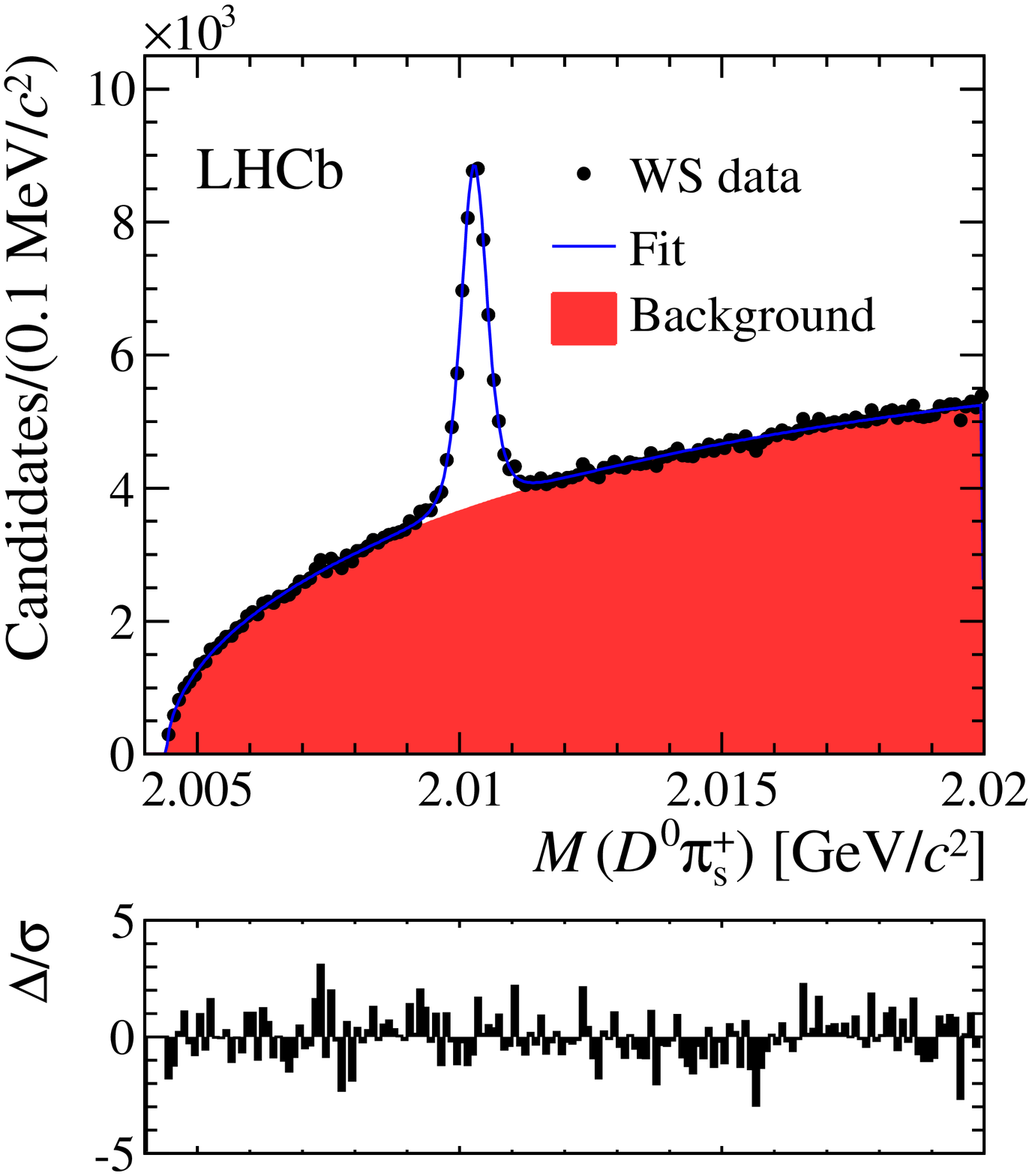}\\
\caption{Time-integrated $\Dz\pis^+$ mass distributions for the selected RS $\Dz\to K^-\pi^+$ (left) and WS $\Dz\to K^+\pi^-$ (right) candidates with fit projections overlaid. The bottom plots show the normalized residuals between the data points and the fits.}\label{fig:yields}
\end{figure*}

\begin{figure}[t]
\centering
\includegraphics[width=0.45\textwidth]{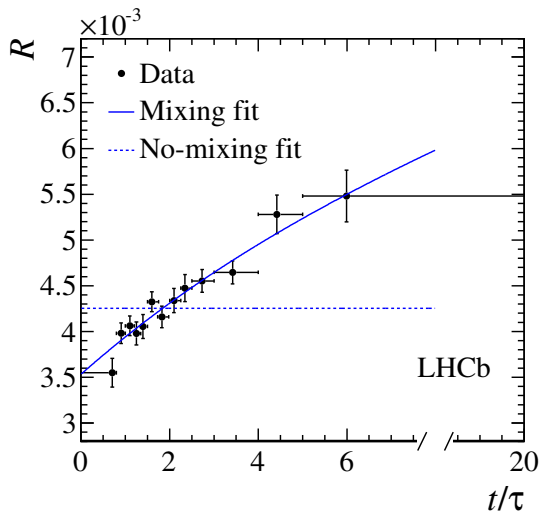}\\
\caption{Decay-time evolution of the ratio, $R$, of WS $\Dz\to K^+\pi^-$ to RS $\Dz\to K^-\pi^+$ yields (points) with the projection of the mixing allowed (solid line) and no-mixing (dashed line) fits overlaid.\label{fig:results}}
\end{figure}

Figure~\ref{fig:yields} shows the \M distribution for the selected RS and WS candidates. Overlaid is the result of a binned $\chi^2$ fit used to separate the \Dstarp signal component, with a mass resolution of about $0.3\,\massmev$, from the background component, which is dominated by associations of real \Dz decays and random pions. The signal mass shape is modeled as the sum of  one Johnson $S_U$ \cite{johnson} and three Gaussian distributions, which account for the asymmetric tails and the central core of the distribution, respectively. The background is described by an empirical function of the form $\left[\M-m_0\right]^ae^{-b\left[\M-m_0\right]}$, where the threshold $m_0$ is fixed to the sum of the known \Dz and $\pi^+$ masses \cite{PDG2012}. We reconstruct approximately $3.6\times10^4$ WS and $8.4\times10^6$ RS decays. To determine the time-dependent WS/RS ratio the data are divided into thirteen \Dz decay time bins, chosen to have a similar number of candidates in each bin. The decay time is estimated from the distance $L$ between the PV and the \Dz decay vertex and from the \Dz momentum as $t/\tau = m_{\Dz} L/p\tau$, where $m_{\Dz}$ and $\tau$ are the known \Dz mass and lifetime \cite{PDG2012}, respectively. The typical decay-time resolution is $\sim0.1\tau$. The signal yields for the RS and WS samples are determined in each decay time bin using fits to the \M distribution. The shape parameters and the yields of the two components, signal and random pion background, are left free to vary in the different decay time bins. We further assume that the \M signal shape for RS and WS decay are the same. Hence, we first perform a fit to the abundant and cleaner RS sample to determine the signal shape and yield, and then, use those shape parameters with fixed values when fitting for the WS signal yield. The signal yields from the thirteen bins are used to calculate the WS/RS ratios, shown in Fig.~\ref{fig:results}, and the mixing parameters are determined in a binned $\chi^2$ fit to the time-dependence according to Eq.~\eqref{eq:true-ratio}.

Since WS and RS events are expected to have the same decay-time acceptance and \M distributions, most systematic uncertainties affecting the determination of the signal yields as a function of decay time cancel in the ratio between WS and RS events. Residual biases from noncanceling instrumental and production effects, such as asymmetries in detection efficiencies or in production, are found to modify the WS/RS ratio only by a relative fraction of $\order(10^{-4})$ and are neglected. Uncertainties in the distance between Vertex Locator sensors can lead to a bias of the decay-time scale. The effect has been estimated to be less than 0.1\% of the measured time \cite{LHCb-PAPER-2012-013} and translates into relative systematic biases of $0.1\%$ and $0.2\%$ on $y'$ and $x'^2$, respectively. At the current level of statistical precision, such small effects are negligible.

The main sources of systematic uncertainty are those which could alter the observed decay-time dependence of the WS/RS ratio. 
Two such sources have been identified: $(1)$ secondary \D mesons, and $(2)$ backgrounds from charm decays reconstructed with the wrong particle identification assignments, which peak in \M and are not accounted for in our mass fit. These effects, discussed below, are expected to depend on the true value of the mixing parameters and are accounted for in the time-dependent fit.

The contamination of charm mesons produced in \bquark-hadron decays could bias the time-dependent measurement, as the reconstructed decay time is calculated with respect to the PV, which, in this case, does not coincide with the \Dz production vertex. When this secondary component is not subtracted, the measured WS/RS ratio can be written as $R(t)\left[1 - \Delta_B(t)\right]$, where $R(t)$ is the ratio of promptly-produced candidates according to Eq.~\eqref{eq:true-ratio}, and $\Delta_B(t)$ is a time-dependent bias due to the secondary contamination. Since $R(t)$ is measured to be monotonically nondecreasing~\cite{hfag} and the reconstructed decay time for secondary decays overestimates the true decay time of the \Dz meson, it is possible to bound $\Delta_B(t)$, for all decay times, as
\begin{equation}\label{eq:Bbias}
0\leqslant \Delta_B(t)\leqslant f_B^{\rm RS}(t)\left[1-\frac{R_D}{R(t)}\right],
\end{equation}
where $f_B^{\rm RS}(t)$ is the fraction of secondary decays in the RS sample at decay time $t$. The lower bound in Eq.~\eqref{eq:Bbias} corresponds to the case when the parent \bquark-hadron decays instantaneously and the reconstructed \Dz decay time is the true decay time. The upper bound corresponds to the case when the \Dz decays instantaneously and the reconstructed decay time $t$ is entirely due to the \bquark-hadron lifetime. Since $\Delta_B\geqslant0$, it follows that the background from secondary $D$ decays decreases the observable mixing effect. To include the corresponding systematic uncertainty, we modify the fitting function for the mixing hypothesis assuming the largest possible bias from Eq.~\eqref{eq:Bbias}. The value of $f_B^{\rm RS}(t)$ is constrained to the measured value, obtained by fitting the $\chi^2(\text{IP})$ distribution of the RS \Dz candidates in bins of decay time. In this fit, the promptly-produced component is described by a time-independent $\chi^2(\text{IP})$ shape, which is derived from data using the candidates with $t<0.8\tau$. The $\chi^2(\text{IP})$ shape of the secondary component, and its dependence on decay time, is also determined from data by studying the sub-sample of candidates that are reconstructed, in combination with other tracks in the events, as $B\to D^*(3)\pi$, $B\to D^*\mu X$ or $B\to D^0\mu X$. The measured value of $f_B^{\rm RS}(t)$ increases almost linearly with decay time from $(0.0\pm0.5)\%$ up to $(14\pm5)\%$, for a time-integrated value of $(2.7\pm0.2)\%$. We checked on pseudoexperiments, before fitting the data, and then also on data that such a small contamination results in a shift on the measured mixing parameters that is much smaller than the increase in the uncertainty when the secondary bias is included in the fit.

Background from incorrectly reconstructed \D meson decays, peaking in the \M distribution, arises from \Dstarp decays for which the correct soft pion is found but the \Dz is partially reconstructed or misidentified. This background is suppressed by the use of tight particle identification and two-body mass requirements. From studies of the events in the \Dz mass sidebands, we find that the dominant peaking background is from RS events that survive the requirements of the WS selection; they are estimated to constitute $(0.4\pm0.2)\%$ of the WS signal. This contamination is expected to have the same decay time dependence of RS decays and, if neglected, would lead to a small increase in the measured value of $R_D$. From the events in the \Dz mass sidebands, we derive a bound on the possible time dependence of this background, which is included in the fit in a similar manner to the secondary background. Contamination from peaking background due to partially reconstructed \Dz decays is found to be much smaller than $0.1\%$ of the WS signal and neglected in the fit.

The $\chi^2$ that is minimized in the fit to the WS/RS decay-time dependence is
\begin{equation}\label{eq:chi2}
\chi^2(r_i,t_i,\sigma_i|\bm{\theta}) = \sum_i\left(\frac{r_i-R(t_i|\bm{\theta})[1-\Delta_B(t_i|\bm{\theta})]-\Delta_p(t_i|\bm{\theta})}{\sigma_i}\right)^2+\chi^2_B(\bm{\theta})+\chi^2_p(\bm{\theta}),
\end{equation}
where $r_i$ and $\sigma_i$ are the measured WS/RS ratio and its statistical uncertainty in the decay time bin $i$, respectively. The decay time $t_i$ is the average value in each bin of the RS sample. The fit parameters, $\bm{\theta}$, include the three mixing parameters ($R_D$, $y'$, $x'^2$) and five nuisance parameters used to describe the decay time evolution of the secondary \D fraction ($\Delta_B$) and of the peaking background ($\Delta_p$). The nuisance parameters are constrained to the measured values by the additional $\chi_B^2$ and $\chi_p^2$ terms, which account for their uncertainties including correlations.

The analysis procedure is defined prior to fitting the data for the mixing parameters. Measurements on pseudoexperiments that mimic the experimental conditions of the data, and where $\Dz-\Dzb$ oscillations are simulated, indicate that the fit procedure is stable and free of any bias.

\begin{table}[t]
\centering
\caption{Results of the time-dependent fit to the data. The uncertainties include statistical and systematic sources; ndf indicates the number of degrees of freedom.
\label{tab:results}}
\begin{tabular}{lcccccc}
\hline\hline
Fit type & Parameter &  $\quad$Fit result  & & \multicolumn{3}{c}{Correlation coefficient}\\
($\chi^2$/ndf) &  & $\quad$($10^{-3}$) & & $R_D$ & $y'$ & $x'^2$ \\
\hline
Mixing & $R_D$ & $\quad3.52\pm0.15$ & & $1$ & $-0.954$ & $+0.882$ \\
($9.5/10$)  & $y'$ & $\quad7.2\pm2.4$ & & & $1$ & $-0.973$ \\
 & $x'^2$ & $\,-0.09\pm0.13$ & & & & 1\\
\hline
No mixing & $R_D$ & $\quad4.25\pm0.04$ & & & &\\
($98.1/12$) & & & & & & \\
\hline\hline
\end{tabular}
\end{table}

\begin{figure}[t]
\centering
\includegraphics[width=0.45\textwidth]{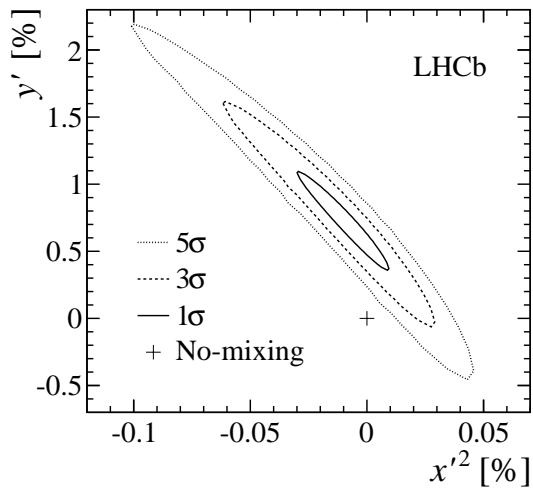}\\
\caption{Estimated confidence-level (CL) regions in the $(x'^2,y')$ plane for $1-\text{CL}=0.317$ ($1\sigma$), $2.7\times10^{-3}$ ($3\sigma$) and $5.73\times10^{-7}$ ($5\sigma$). Systematic uncertainties are included. The cross indicates the no-mixing point.}\label{fig:contours}
\end{figure}

The fit to the decay-time evolution of the WS/RS ratio is shown in Fig.~\ref{fig:results} (solid line), with the values and uncertainties of the parameters $R_D$, $y'$ and $x'^2$ listed in Table~\ref{tab:results}. The value of $x'^2$ is found to be negative, but consistent with zero. As the dominant systematic uncertainties are treated within the fit procedure (all other systematic effects are negligible), the quoted errors account for systematic as well as statistical uncertainties. When the systematic biases are not included in the fit, the estimated uncertainties on $R_D$, $y'$ and $x'^2$ become respectively $6\%$, $10\%$ and $11\%$ smaller, showing that the quoted uncertainties are dominated by their statistical component. To evaluate the significance of this mixing result we determine the change in the fit $\chi^2$ when the data are described under the assumption of the no-mixing hypothesis (dashed line in Fig.~\ref{fig:results}). Under the assumption that the $\chi^2$ difference, $\Delta \chi^2$, follows a $\chi^2$ distribution for two degrees of freedom, $\Delta\chi^2=88.6$ corresponds to a $p$-value of $5.7\times 10^{-20}$, which excludes the no-mixing hypothesis at $9.1$ standard deviations. This is illustrated in Fig.~\ref{fig:contours} where the $1\sigma$, $3\sigma$ and $5\sigma$ confidence regions for $x'^2$ and $y'$ are shown.

As additional cross-checks, we perform the measurement in statistically independent sub-samples of the data, selected according to different data-taking periods, and find compatible results. We also use alternative decay-time binning schemes, selection criteria or fit methods to separate signal and background, and find no significant variations in the estimated parameters. Finally, to assess the impact of events where more than one candidate is reconstructed, we repeat the time-dependent fit on data after randomly removing the additional candidates and selecting only one per event; the change in the measured value of $R_D$, $y'$ and $x'^2$ is $2\%$, $6\%$ and $7\%$ of their uncertainty, respectively.

In conclusion, we measure the decay time dependence of the ratio between $\Dz\to K^+\pi^-$ and $\Dz\to K^-\pi^+$ decays using $1.0\invfb$ of data and exclude the no-mixing hypothesis at $9.1$ standard deviations. This is the first observation of $\Dz-\Dzb$ oscillations in a single measurement. The measured values of the mixing parameters are compatible with and have substantially better precision than those from previous measurements \cite{Aubert:2007wf,Aaltonen:2007ac,Zhang:2006dp}.

\section*{Acknowledgements}
\noindent This Letter is dedicated to the memory of our friend and colleague Javier Magnin.
We express our gratitude to our colleagues in the CERN
accelerator departments for the excellent performance of the LHC. We
thank the technical and administrative staff at the LHCb
institutes. We acknowledge support from CERN and from the national
agencies: CAPES, CNPq, FAPERJ and FINEP (Brazil); NSFC (China);
CNRS/IN2P3 and Region Auvergne (France); BMBF, DFG, HGF and MPG
(Germany); SFI (Ireland); INFN (Italy); FOM and NWO (The Netherlands);
SCSR (Poland); ANCS/IFA (Romania); MinES, Rosatom, RFBR and NRC
``Kurchatov Institute'' (Russia); MinECo, XuntaGal and GENCAT (Spain);
SNSF and SER (Switzerland); NAS Ukraine (Ukraine); STFC (United
Kingdom); NSF (USA). We also acknowledge the support received from the
ERC under FP7. The Tier1 computing centres are supported by IN2P3
(France), KIT and BMBF (Germany), INFN (Italy), NWO and SURF (The
Netherlands), PIC (Spain), GridPP (United Kingdom). We are thankful
for the computing resources put at our disposal by Yandex LLC
(Russia), as well as to the communities behind the multiple open
source software packages that we depend on.

\bibliographystyle{LHCb}
\bibliography{main}

\ifx\mcitethebibliography\mciteundefinedmacro
\PackageError{LHCb.bst}{mciteplus.sty has not been loaded}
{This bibstyle requires the use of the mciteplus package.}\fi
\providecommand{\href}[2]{#2}
\begin{mcitethebibliography}{10}
\mciteSetBstSublistMode{n}
\mciteSetBstMaxWidthForm{subitem}{\alph{mcitesubitemcount})}
\mciteSetBstSublistLabelBeginEnd{\mcitemaxwidthsubitemform\space}
{\relax}{\relax}

\bibitem{Lande:1956pf}
K.~Lande {\em et~al.}, \ifthenelse{\boolean{articletitles}}{{\it {Observation
  of long-lived neutral $V$ particles}},
  }{}\href{http://dx.doi.org/10.1103/PhysRev.103.1901}{Phys.\ Rev.\  {\bf 103}
  (1956) 1901}\relax
\mciteBstWouldAddEndPuncttrue
\mciteSetBstMidEndSepPunct{\mcitedefaultmidpunct}
{\mcitedefaultendpunct}{\mcitedefaultseppunct}\relax
\EndOfBibitem
\bibitem{Albrecht:1987dr}
ARGUS collaboration, H.~Albrecht {\em et~al.},
  \ifthenelse{\boolean{articletitles}}{{\it {Observation of $\Bz-\Bzb$
  mixing}}, }{}\href{http://dx.doi.org/10.1016/0370-2693(87)91177-4}{Phys.\
  Lett.\  {\bf B 192} (1987) 245}\relax
\mciteBstWouldAddEndPuncttrue
\mciteSetBstMidEndSepPunct{\mcitedefaultmidpunct}
{\mcitedefaultendpunct}{\mcitedefaultseppunct}\relax
\EndOfBibitem
\bibitem{Abulencia:2006ze}
CDF collaboration, A.~Abulencia {\em et~al.},
  \ifthenelse{\boolean{articletitles}}{{\it {Observation of $\Bz_{\squark}
  -\Bzb_{\squark}$ oscillations}},
  }{}\href{http://dx.doi.org/10.1103/PhysRevLett.97.242003}{Phys.\ Rev.\ Lett.\
   {\bf 97} (2006) 242003}, \href{http://arxiv.org/abs/hep-ex/0609040}{{\tt
  arXiv:hep-ex/0609040}}\relax
\mciteBstWouldAddEndPuncttrue
\mciteSetBstMidEndSepPunct{\mcitedefaultmidpunct}
{\mcitedefaultendpunct}{\mcitedefaultseppunct}\relax
\EndOfBibitem
\bibitem{Aubert:2007wf}
\babar collaboration, B.~Aubert {\em et~al.},
  \ifthenelse{\boolean{articletitles}}{{\it {Evidence for $\Dz-\Dzb$ mixing}},
  }{}\href{http://dx.doi.org/10.1103/PhysRevLett.98.211802}{Phys.\ Rev.\ Lett.\
   {\bf 98} (2007) 211802}, \href{http://arxiv.org/abs/hep-ex/0703020}{{\tt
  arXiv:hep-ex/0703020}}\relax
\mciteBstWouldAddEndPuncttrue
\mciteSetBstMidEndSepPunct{\mcitedefaultmidpunct}
{\mcitedefaultendpunct}{\mcitedefaultseppunct}\relax
\EndOfBibitem
\bibitem{Staric:2007dt}
Belle collaboration, M.~Staric {\em et~al.},
  \ifthenelse{\boolean{articletitles}}{{\it {Evidence for $\Dz-\Dzb$ mixing}},
  }{}\href{http://dx.doi.org/10.1103/PhysRevLett.98.211803}{Phys.\ Rev.\ Lett.\
   {\bf 98} (2007) 211803}, \href{http://arxiv.org/abs/hep-ex/0703036}{{\tt
  arXiv:hep-ex/0703036}}\relax
\mciteBstWouldAddEndPuncttrue
\mciteSetBstMidEndSepPunct{\mcitedefaultmidpunct}
{\mcitedefaultendpunct}{\mcitedefaultseppunct}\relax
\EndOfBibitem
\bibitem{Aaltonen:2007ac}
CDF collaboration, T.~Aaltonen {\em et~al.},
  \ifthenelse{\boolean{articletitles}}{{\it {Evidence for $\Dz-\Dzb$ mixing
  using the CDF II detector}},
  }{}\href{http://dx.doi.org/10.1103/PhysRevLett.100.121802}{Phys.\ Rev.\
  Lett.\  {\bf 100} (2008) 121802}, \href{http://arxiv.org/abs/0712.1567}{{\tt
  arXiv:0712.1567}}\relax
\mciteBstWouldAddEndPuncttrue
\mciteSetBstMidEndSepPunct{\mcitedefaultmidpunct}
{\mcitedefaultendpunct}{\mcitedefaultseppunct}\relax
\EndOfBibitem
\bibitem{Aubert:2008zh}
\babar collaboration, B.~Aubert {\em et~al.},
  \ifthenelse{\boolean{articletitles}}{{\it {Measurement of $D^0 -
  \overline{D}^0$ mixing from a time-dependent amplitude analysis of $D^0 \to
  K^{+} \pi^{-} \pi^0$ decays}},
  }{}\href{http://dx.doi.org/10.1103/PhysRevLett.103.211801}{Phys.\ Rev.\
  Lett.\  {\bf 103} (2009) 211801}, \href{http://arxiv.org/abs/0807.4544}{{\tt
  arXiv:0807.4544}}\relax
\mciteBstWouldAddEndPuncttrue
\mciteSetBstMidEndSepPunct{\mcitedefaultmidpunct}
{\mcitedefaultendpunct}{\mcitedefaultseppunct}\relax
\EndOfBibitem
\bibitem{Aubert:2009ai}
\babar collaboration, B.~Aubert {\em et~al.},
  \ifthenelse{\boolean{articletitles}}{{\it {Measurement of $\Dz-\Dzb$ mixing
  using the ratio of lifetimes for the decays $D^0\to K^-\pi^+$ and $K^+K^-$}},
  }{}\href{http://dx.doi.org/10.1103/PhysRevD.80.071103}{Phys.\ Rev.\  {\bf D
  80} (2009) 071103}, \href{http://arxiv.org/abs/0908.0761}{{\tt
  arXiv:0908.0761}}\relax
\mciteBstWouldAddEndPuncttrue
\mciteSetBstMidEndSepPunct{\mcitedefaultmidpunct}
{\mcitedefaultendpunct}{\mcitedefaultseppunct}\relax
\EndOfBibitem
\bibitem{hfag}
Heavy Flavor Averaging Group, Y.~Amhis {\em et~al.},
  \ifthenelse{\boolean{articletitles}}{{\it {Averages of $b$-hadron,
  $c$-hadron, and $\tau$-lepton properties as of early 2012}},
  }{}\href{http://arxiv.org/abs/1207.1158}{{\tt arXiv:1207.1158}}\relax
\mciteBstWouldAddEndPuncttrue
\mciteSetBstMidEndSepPunct{\mcitedefaultmidpunct}
{\mcitedefaultendpunct}{\mcitedefaultseppunct}\relax
\EndOfBibitem
\bibitem{Bianco:2003vb}
S.~Bianco, F.~Fabbri, D.~Benson, and I.~Bigi,
  \ifthenelse{\boolean{articletitles}}{{\it {A Cicerone for the physics of
  charm}}, }{}Riv.\ Nuovo Cim.\  {\bf 26N7} (2003) 1,
  \href{http://arxiv.org/abs/hep-ex/0309021}{{\tt arXiv:hep-ex/0309021}}\relax
\mciteBstWouldAddEndPuncttrue
\mciteSetBstMidEndSepPunct{\mcitedefaultmidpunct}
{\mcitedefaultendpunct}{\mcitedefaultseppunct}\relax
\EndOfBibitem
\bibitem{Burdman:2003rs}
G.~Burdman and I.~Shipsey, \ifthenelse{\boolean{articletitles}}{{\it
  {$\Dz-\Dzb$ mixing and rare charm decays}},
  }{}\href{http://dx.doi.org/10.1146/annurev.nucl.53.041002.110348}{Ann.\ Rev.\
  Nucl.\ Part.\ Sci.\  {\bf 53} (2003) 431},
  \href{http://arxiv.org/abs/hep-ph/0310076}{{\tt arXiv:hep-ph/0310076}}\relax
\mciteBstWouldAddEndPuncttrue
\mciteSetBstMidEndSepPunct{\mcitedefaultmidpunct}
{\mcitedefaultendpunct}{\mcitedefaultseppunct}\relax
\EndOfBibitem
\bibitem{Shipsey:2006zz}
I.~Shipsey, \ifthenelse{\boolean{articletitles}}{{\it {Status of charm flavor
  physics}}, }{}\href{http://dx.doi.org/10.1142/S0217751X06034525}{Int.\ J.\
  Mod.\ Phys.\  {\bf A 21} (2006) 5381},
  \href{http://arxiv.org/abs/hep-ex/0607070}{{\tt arXiv:hep-ex/0607070}}\relax
\mciteBstWouldAddEndPuncttrue
\mciteSetBstMidEndSepPunct{\mcitedefaultmidpunct}
{\mcitedefaultendpunct}{\mcitedefaultseppunct}\relax
\EndOfBibitem
\bibitem{Artuso:2008vf}
M.~Artuso, B.~Meadows, and A.~A. Petrov,
  \ifthenelse{\boolean{articletitles}}{{\it {Charm meson decays}},
  }{}\href{http://dx.doi.org/10.1146/annurev.nucl.58.110707.171131}{Ann.\ Rev.\
  Nucl.\ Part.\ Sci.\  {\bf 58} (2008) 249},
  \href{http://arxiv.org/abs/0802.2934}{{\tt arXiv:0802.2934}}\relax
\mciteBstWouldAddEndPuncttrue
\mciteSetBstMidEndSepPunct{\mcitedefaultmidpunct}
{\mcitedefaultendpunct}{\mcitedefaultseppunct}\relax
\EndOfBibitem
\bibitem{Petrov:2006nc}
A.~A. Petrov, \ifthenelse{\boolean{articletitles}}{{\it {Charm mixing in the
  standard model and beyond}},
  }{}\href{http://dx.doi.org/10.1142/S0217751X06034902}{Int.\ J.\ Mod.\ Phys.\
  {\bf A 21} (2006) 5686}, \href{http://arxiv.org/abs/hep-ph/0611361}{{\tt
  arXiv:hep-ph/0611361}}\relax
\mciteBstWouldAddEndPuncttrue
\mciteSetBstMidEndSepPunct{\mcitedefaultmidpunct}
{\mcitedefaultendpunct}{\mcitedefaultseppunct}\relax
\EndOfBibitem
\bibitem{Golowich:2007ka}
E.~Golowich, J.~Hewett, S.~Pakvasa, and A.~A. Petrov,
  \ifthenelse{\boolean{articletitles}}{{\it {Implications of $\Dz-\Dzb$ mixing
  for new physics}},
  }{}\href{http://dx.doi.org/10.1103/PhysRevD.76.095009}{Phys.\ Rev.\  {\bf D
  76} (2007) 095009}, \href{http://arxiv.org/abs/0705.3650}{{\tt
  arXiv:0705.3650}}\relax
\mciteBstWouldAddEndPuncttrue
\mciteSetBstMidEndSepPunct{\mcitedefaultmidpunct}
{\mcitedefaultendpunct}{\mcitedefaultseppunct}\relax
\EndOfBibitem
\bibitem{Ciuchini:2007cw}
M.~Ciuchini {\em et~al.}, \ifthenelse{\boolean{articletitles}}{{\it {$\Dz-\Dzb$
  mixing and new physics: general considerations and constraints on the MSSM}},
  }{}\href{http://dx.doi.org/10.1016/j.physletb.2007.08.055}{Phys.\ Lett.\
  {\bf B 655} (2007) 162}, \href{http://arxiv.org/abs/hep-ph/0703204}{{\tt
  arXiv:hep-ph/0703204}}\relax
\mciteBstWouldAddEndPuncttrue
\mciteSetBstMidEndSepPunct{\mcitedefaultmidpunct}
{\mcitedefaultendpunct}{\mcitedefaultseppunct}\relax
\EndOfBibitem
\bibitem{Alves:2008zz}
LHCb collaboration, A.~A. Alves~Jr. {\em et~al.},
  \ifthenelse{\boolean{articletitles}}{{\it {The \lhcb detector at the LHC}},
  }{}\href{http://dx.doi.org/10.1088/1748-0221/3/08/S08005}{JINST {\bf 3}
  (2008) S08005}\relax
\mciteBstWouldAddEndPuncttrue
\mciteSetBstMidEndSepPunct{\mcitedefaultmidpunct}
{\mcitedefaultendpunct}{\mcitedefaultseppunct}\relax
\EndOfBibitem
\bibitem{PDG2012}
Particle Data Group, J.~Beringer {\em et~al.},
  \ifthenelse{\boolean{articletitles}}{{\it {\href{http://pdg.lbl.gov/}{Review
  of particle physics}}}, }{}Phys.\ Rev.\  {\bf D 86} (2012) 010001\relax
\mciteBstWouldAddEndPuncttrue
\mciteSetBstMidEndSepPunct{\mcitedefaultmidpunct}
{\mcitedefaultendpunct}{\mcitedefaultseppunct}\relax
\EndOfBibitem
\bibitem{Aaltonen:2011se}
CDF collaboration, T.~Aaltonen {\em et~al.},
  \ifthenelse{\boolean{articletitles}}{{\it {Measurement of CP--violating
  asymmetries in $D^0\to\pi^+\pi^-$ and $D^0\to K^+K^-$ decays at CDF}},
  }{}\href{http://dx.doi.org/10.1103/PhysRevD.85.012009}{Phys.\ Rev.\  {\bf D
  85} (2012) 012009}, \href{http://arxiv.org/abs/1111.5023}{{\tt
  arXiv:1111.5023}}\relax
\mciteBstWouldAddEndPuncttrue
\mciteSetBstMidEndSepPunct{\mcitedefaultmidpunct}
{\mcitedefaultendpunct}{\mcitedefaultseppunct}\relax
\EndOfBibitem
\bibitem{johnson}
N.~L. Johnson, \ifthenelse{\boolean{articletitles}}{{\it {Systems of frequency
  curves generated by methods of translation}}, }{}Biometrika {\bf 36} (1949)
  149\relax
\mciteBstWouldAddEndPuncttrue
\mciteSetBstMidEndSepPunct{\mcitedefaultmidpunct}
{\mcitedefaultendpunct}{\mcitedefaultseppunct}\relax
\EndOfBibitem
\bibitem{LHCb-PAPER-2012-013}
LHCb collaboration, R.~Aaij {\em et~al.},
  \ifthenelse{\boolean{articletitles}}{{\it {Measurement of the effective
  $B_s^0 \rightarrow K^+ K^-$ lifetime}},
  }{}\href{http://dx.doi.org/10.1016/j.physletb.2012.08.033}{Phys.\ Lett.\
  {\bf B 716} (2012) 393}, \href{http://arxiv.org/abs/1207.5993}{{\tt
  arXiv:1207.5993}}\relax
\mciteBstWouldAddEndPuncttrue
\mciteSetBstMidEndSepPunct{\mcitedefaultmidpunct}
{\mcitedefaultendpunct}{\mcitedefaultseppunct}\relax
\EndOfBibitem
\bibitem{Zhang:2006dp}
Belle collaboration, L.~Zhang {\em et~al.},
  \ifthenelse{\boolean{articletitles}}{{\it {Improved constraints on $\Dz-\Dzb$
  mixing in $\Dz\to K^+\pi^-$ decays at Belle}},
  }{}\href{http://dx.doi.org/10.1103/PhysRevLett.96.151801}{Phys.\ Rev.\ Lett.\
   {\bf 96} (2006) 151801}, \href{http://arxiv.org/abs/hep-ex/0601029}{{\tt
  arXiv:hep-ex/0601029}}\relax
\mciteBstWouldAddEndPuncttrue
\mciteSetBstMidEndSepPunct{\mcitedefaultmidpunct}
{\mcitedefaultendpunct}{\mcitedefaultseppunct}\relax
\EndOfBibitem
\end{mcitethebibliography}

\end{document}